# Slow dynamic nonlinearity in unconsolidated glass bead packs

John Y. Yoritomo and Richard L. Weaver
Department of Physics
University of Illinois, Urbana, IL 61801

**Abstract**

Slow dynamic nonlinearity describes a poorly understood, creep-like phenomena that occurs in brittle composite materials such as rocks and cement. It is characterized by a drop in stiffness induced by a mechanical conditioning, followed by a log(time) recovery. A consensus theoretical understanding of the behavior has not been developed. Here we introduce an alternative experimental venue with which to inform theory. Unconsolidated glass bead packs are studied rather than rocks or cement because the structure and internal contacts of bead packs are less complex and better understood. Slow dynamics has been observed in such systems previously. However, the measurements to date tend to be irregular. Particular care is used here in the experimental design to overcome the difficulties inherent in bead pack studies. This includes the design of the bead pack support, the use of low frequency conditioning, and the use of ultrasonic waves as a probe with coda wave interferometry to assess changes. Slow dynamics is observed in our system after three different methods for low-frequency conditioning, one of which has not been reported in the literature previously.

## I. Introduction

A wide range of solid materials demonstrate fascinating loss of stiffness after a mechanically-induced conditioning ("pumping"), followed by gradual log(time) recovery. These behaviors are known as slow dynamic nonlinearity and appear to be universal amongst porous ceramic and granular materials such as concretes and natural rocks. In spite of the ubiquity of these behaviors, they are poorly understood.

Studies of non-classical nonlinear elasticity in rocks and cement-based materials, led by remarkable work at Los Alamos National Lab (LANL) [1–7] and employing Nonlinear Resonant Ultrasound Spectroscopy (NRUS), have found that the application of minor conditioning strain (as little as $10^{-6}$) leads to a drop in elastic modulus. The loss of stiffness is followed, after the strain is removed, by a slow recovery towards the original value. In much of the work, the conditioning strain (or "pumping") was applied through oscillatory vibrations at frequencies of a few kHz. Loss of stiffness induced by the vibration was revealed in the sample's decreased fundamental vibration frequencies (also at a few kHz). But more significantly, the modulus was found to heal over a period from a few seconds to hours after the conditioning strain was removed, where healing progressed with the logarithm of time since the conditioning ended. Similar behavior was observed in material "Q" related to absorption of mechanical energy. The same behaviors were observed after conditioning by temperature and humidity changes [8]. The behaviors are observed to scale with the first power of conditioning strain amplitude [2]. Neither the recoveries, nor their time dependence, are well understood.

Loss of stiffness and log(t) recovery are also seen in seismic (at ~1 Hz) wave speed near a fault after an earthquake, e.g. refs. [9,10], where recoveries were monitored over periods from days to years and correlated with aftershocks. This behavior is not well understood either.

Other laboratory techniques have been used to monitor changes and recoveries. Lobkis and Weaver [11] monitored slow dynamic recoveries of narrow-band ultrasonic Larsen frequency in sandstone and cement paste samples after impact conditioning, where they were able to detect log(t) like recovery as soon as three milliseconds after the impact. Tremblay *et al.* [12], after conditioning by impacts in concrete, monitored broad-band diffuse reverberant ultrasonic signals and measured changes using coda wave interferometry. Shokouhi *et al.* [7], after pumping with fundamental frequency vibrations, used Dynamic Acousto-Elasticity Testing (DAET) [13,14] to monitor changes and recoveries in the transit time of a high frequency ultrasonic pulse. Shokouhi *et al.* [7] fit the observed relaxations to a discrete sum of exponential relaxations.

Related work (e.g., ref. [15,16]) has focused on fast nonlinear dynamics, in which the peculiar nonlinearity of these materials is examined at the finest time resolutions, comparable to or within the period of the pump conditioning.

It is widely supposed that the unusual nonlinearity of rocks in general and slow dynamics in particular have their origin in the glassy contacts between crystallites and to the breaking and healing of bonds or joins there, like that seen in dry friction [17]. But beyond that there is little consensus, in particular in regard to the mechanism of the recovery or the nature of the bonds. TenCate *et al.* [18], noted that the slowness might be due to a distribution of activation energies (uniform between 0.5 to 1.0 eV) associated with atomic-scale barriers that are overcome by thermal fluctuations. They showed that this model implies that the log(t) relaxation should proceed at a rate proportional to temperature. Their attempts to measure such dependence were inconclusive. Others have proposed similar models [19–22]. All require the distribution to be uniform over some range. No satisfactory explanation of the physical mechanism behind this uniformity—or why it is so universal—has been given.

Moisture has been suggested as relevant [23–26]. Bocquet *et al.* [27] discuss moisture-induced aging for friction in granular media and derive a humidity-dependent log(t) behavior governed by a thermal activation process. Bittner [25] showed that fully saturated cements did not exhibit slow dynamics and has suggested that diffusion of water vapor along cracks is responsible for the slowness. However, TenCate's [3] studies of an almost fully dried sandstone sample held in vacuum for months did not show loss of slow dynamics, suggesting water is not responsible.

Models include that of Vakhnenko *et al.* [28], who offer a soft-rachet model of ruptured and recovering intergrain defects that reproduces log(t) like slow dynamics, and Snieder *et al.* [20], who show that they can fit a log(t) relaxation to a distribution of exponential relaxations. Attempts have been made to explain how slow dynamics is related to other nonlinear phenomena associated with rocks as well. Zaitsev *et al.* [29] suggest an origin for both hysteresis and slow dynamics in bistable contacts between grains. A similar suggestion is made by Lebedev and Ostrovsky [21], whose model incorporates two types of contact forces, Herztian-elastic and adhesion, and a metastable state (see also [22]). The (assumed uniform distribution of) activation energy model of Li *et al.* [19] reproduces log(t) slow dynamic recovery as well as certain observed sweep-rate dependences in resonance curves.

In spite of the many phenomenological fits and some plausible hypotheses, there is still little consensus on the micro-physics ultimately responsible for slow dynamics' remarkable log(t) linearity or ubiquity. This may in part be because the microstructures are so poorly understood. Rocks and cements are highly complex multi-phase materials, in general consisting of water, crystallites, cracks, inclusions, glassy contacts, residual stresses, and slow chemical reactions.

Studies in simpler structures may therefore be of value. Zaitsev *et al.* [30] demonstrated slow dynamics in glass rods with a small number of thermal cracks. (They also suggest a thermoelastic mechanism in which the slowness is owed to thermal diffusion around microcracks.) Slow dynamics has also been observed in other cracked glass structures by Johnson and Sutin [4] and Bittner [25]. Both report no slow dynamics in pristine, crack-less glass bodies.

It may be argued that unconsolidated glass bead assemblages are even simpler than the glass systems mentioned above. The structure and internal contacts of bead packs are better understood than the crack geometries of the glasses. Depending on pore size the packs may also allow ready and controlled ingress of heat and water vapor. Slow dynamics has been observed in such systems. Johnson and Jia [31] present evidence of slow dynamics using NRUS at 17kHz. Their recoveries, while highly irregular, appeared logarithmic from minutes to hours. Tournat and Gusev [32] focused on bead pack acoustic nonlinearity in general rather than just slow dynamics, but show evidence for it in passing. They introduce a "resemblance parameter" similar to the distortion parameter of coda wave interferometry [33] with which they quantify how their high frequency diffuse waveform varies during the pump phase and the relaxation. Slow relaxation of the amplitude of a nonlinearly demodulated wave [29] and of a nonlinearly-induced modulation sidelobe [34] have been demonstrated in glass beads as well. Jia *et al.* [35] demonstrate slow dynamic log(t) recovery of the low frequency ballistic ultrasound speed, increasing from an (enormous by the standards of the field) 8% deficit relative to the base speed to a 4% deficit after 8000 seconds.

Unfortunately, the measurements to date have tended to be irregular and contain a high degree of noise. One challenge is loading and supporting the unconsolidated packs without contaminating the acoustics. Glass bead packs are further complicated by their complex albeit fascinating acoustics [32,36]. Nonlinearity is strong, especially at low static confining pressures. Even the linear regime is complex; high frequency waves are strongly scattered and highly diffuse [37–39].

Here we present our studies of ultrasound in glass bead packs and employ coda wave interferometry as a probe of slow dynamic recoveries after various conditionings. We suggest this system could be a useful venue for examining the effects of various structural parameters on slow dynamics and thereby informing theory for its microphysical basis. In the next section we describe the experimental design, and the propagation and spectrum of linear ultrasound in the bead pack. The subsequent section presents the coda wave interferometry technique [33,40,41] for measuring tiny changes in diffuse ultrasonic waveforms. Then we present the results of the slow dynamic experiments and conclude with a discussion of the advantages of this system for slow dynamic studies.

## II. Experimental Design and Ultrasound Propagation

The experimental design is shown in Fig. 1. Common soda-lime glass beads, nominally mono-disperse with a diameter of 2.97 $\pm$ 0.05mm and a mass of 30mg, were used to construct a cylindrical bead pack 71mm in diameter and 33mm thick. The mass of the bead pack is measured to be 221g, corresponding to a density of $\rho_{bp} = 1.69 g/cm^3$. The packing fraction, $\eta = \rho_{bp}/\rho_{glass}$ is 0.67 (using $\rho_{glass} = 2.52$), much less than hexagonal close-packed (0.74) but greater than random close-packed (0.64).[1] The pack is sandwiched by 1.6mm steel plates, which are in turn sandwiched by high strength foam (FOAMULAR® 1000 Extruded Polystyrene (XPS), with a quoted strength of 100 psi). The bead pack is surrounded on the sides by the same high strength foam. To ensure uniform force distribution in the bead pack [42], the foams walls are floating, i.e. they are held up solely from the frictional force with the beads. A 87kg (or 215kPa) dead-weight static load consisting of four steel slabs (label (i) in Fig. 1b) is placed on top of the bead pack. This load assures maximum coordination number and good contact between the beads [36]; further load, and in particular conditioning strains, will not change topology. It also leads to the amplitude of ultrasound we use being in the linear regime. We prefer a dead-weight static load—as opposed to an active press—because we can accurately estimate the pressure on the bead pack and have no interference from noise associated with the press. After the beads are assembled and stirred, the structure is shaken to encourage settling of the beads. Also a few cycles of adding and removing the steel slabs are performed. We allow the bead pack to settle for at least 24 hours before a measurement is conducted.

As the 87kg load is resting on a comparatively small area, four legs are used for a safety precaution against toppling (label (ii) in Fig. 1b). Rubber shims were put between the tops of these legs and the static load (label (iii) in Fig. 1b). These were used to reduce swaying from the conditioning as well as background vibrations (20-30Hz) of the laboratory floor. The vibrations noisily modulated the ultrasonic signals, which in turn increased the noise of the coda wave interferometry analysis (next section). We believe the legs to take up a minimal amount of the 87kg weight, however, and we have verified that the rubber shims do not affect the slow dynamic measurements (Sec. 4), except in that they reduce the noise.

The source and receiver ultrasonic transducers (Physical Acoustics Corp. (Mistras) micro30) are coupled to the top and bottom steel plates with glue. A 10ns-duration high-voltage broadband ultrasonic pulse is sent to the upper transducer every 0.01 seconds. The received signal at the lower transducer is amplified by a 40dB ultrasonic preamplifier (Panametrics model 5670) and then recorded by a digitizer (GaGe CSE8442) at 10Msamples/sec. 100 received signals are repetition averaged to improve signal-to-noise. A repetition-averaged ultrasonic signal is produced approximately every three seconds (the acquisition software consumes two seconds). Changes in these signals are quantified using coda wave interferometry (next section); these changes are used to assess how the bead pack responds to conditioning (section IV).

---

[1] The difference from rpc is ascribable to surface effects and residual crystalline order near the steel plates.

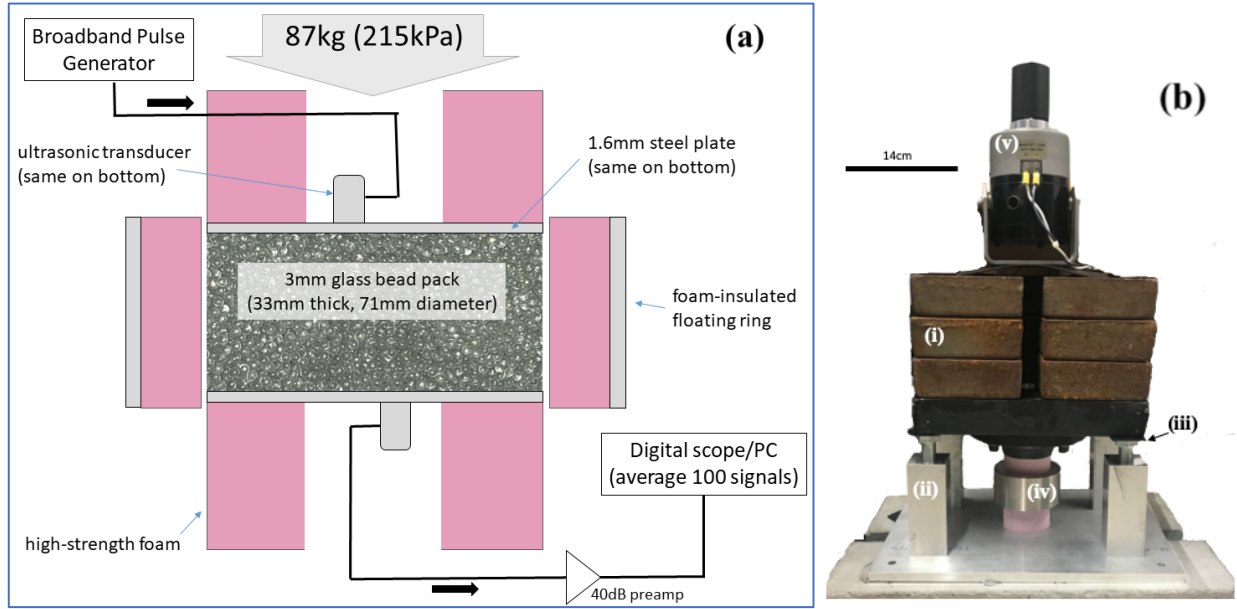

Figure 1. A schematic, panel (a), and photo, panel (b), of the experimental setup. Panel (a) shows that the bead pack is sandwiched by 1.6mm steel plates, which, in turn, are sandwiched by thick-walled hollow cylinders of high strength foam. The same foam is used to confine the beads laterally. A broadband pulse is sent to the top steel plate where it spreads laterally and then propagates into the bead pack. A transducer at the bottom receives the signal, which is then sent to a preamp before being digitized. The foam walls are themselves surrounded by a metal ring to support the foam walls against splitting. The walls are floating; nothing is supporting them from the bottom, but rather they are held up solely by the frictional forces with the beads. This is to ensure a uniform force distribution [42]. The picture in panel (b) shows the four 5 x 30.5 x 21cm steel slabs (i) that rest on top of the bead pack (iv). They result in a 215kPa static pressure on the bead pack. The four legs (ii) near the bottom are meant only for safety, as a precaution against toppling. Rubber shims (iii) were loosely placed between the tops of these legs and the static load. These were used to reduce swaying from the conditioning and from background vibrations (20-30Hz) of the laboratory floor. We believe the legs to take up a minimal amount of the 87kg weight, however, and we have verified that the rubber shims do not affect the slow dynamic experiments. The picture also shows the dynamical shaker (v) used in harmonic conditioning (Sec. IV). It rests on top of the static load and drives a 1kg mass (black object at very top) at 60Hz.

Figure 2 shows a typical ultrasonic signal (Fig. 2a) through the bead pack and its spectrum (blue curve in Fig. 2b). The signal lasts a couple of milliseconds. A first arrival can be identified at 40 microseconds (inset of Fig. 2a). It is notably lower frequency (~65kHz) than what arrives later (~150kHz). The first arrival time corresponds to a wavespeed of $c = 825 m/s$, comparable to other studies of ultrasound in unconsolidated glass bead packs [39]. Most of the energy arrives later, at around 400 microseconds, suggesting diffusive transport. The energy can be fit to a diffusion equation (Figure 3) with a diffusivity of $D = 0.36 m^2/s$ and an absorption of $\alpha = 4200 \ s^{-1}$. This diffusivity is comparable to that reported by Jia [39], when rescaled for differences in bead diameter and central frequency and with allowances for different pressure. This diffusivity corresponds to a dwell time $\tau_{dwell} = a^2/(6D) = 4 \mu s$ (where $a = 3mm$ is the diameter of a bead), obtained from the diffusivity of a 3-d random walk with random steps $a$ every $\tau_{dwell}$.

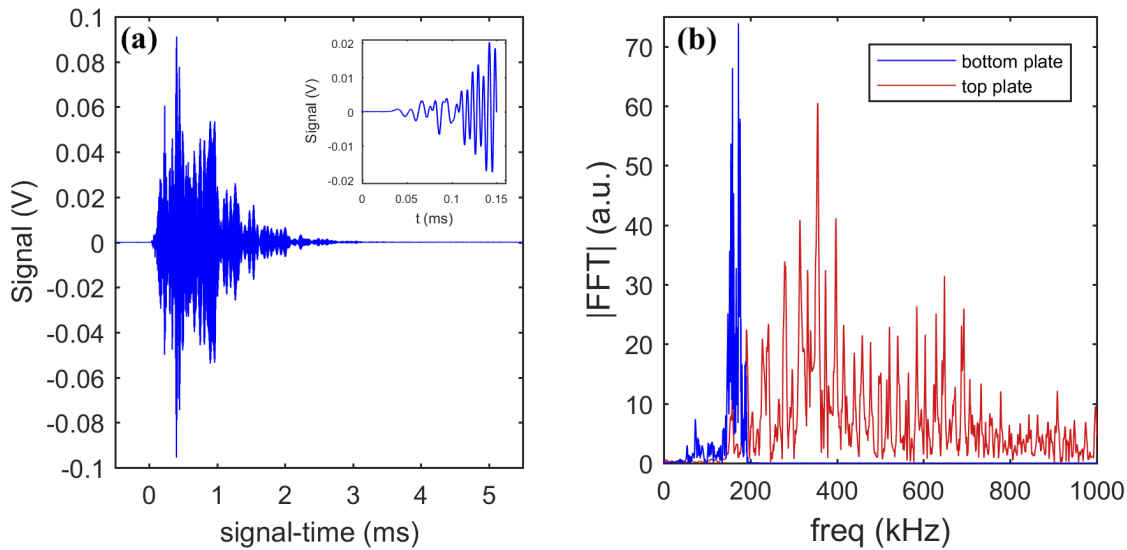

Figure 2. Typical ultrasonic signal (2a) and its spectrum (blue curve in 2b) through the glass bead pack. The inset in panel (a) shows the signal at very early times. A first arrival at 40 microseconds can be identified that is notably lower frequency (~65kHz) than what arrives after 100µ$s$. Most of energy arrives even later, suggesting diffusive transport. The spectrum of the signal through the bead pack is the blue curve in panel (b). It has a sharp cut-off frequency near 200kHz. The bead pack acts as a low pass filter because much higher frequencies are sent into the bead pack, as evidenced by the red curve in panel (b).

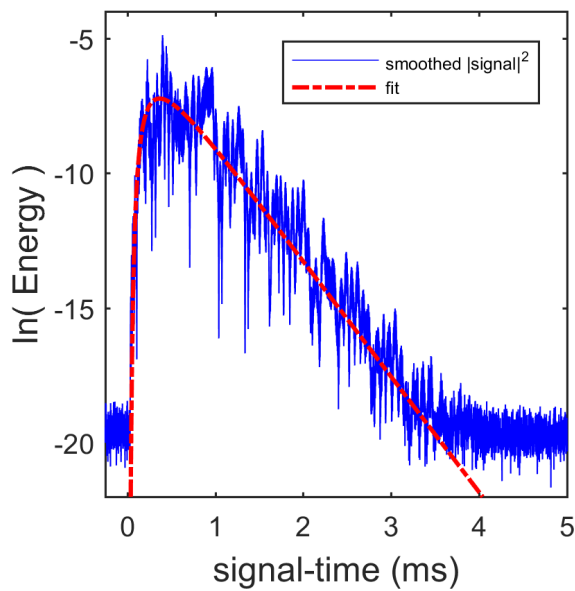

Figure 3. A plot of energy (smoothed signal squared) versus signal-time (solid blue curve). A fit to a diffusion equation is plotted as well (dashed red curve): $E = \frac{A}{\sqrt{t}}e^{-\alpha t - L^2/4Dt}$. A diffusivity of $D = 0.36 m^2/s$ and an absorption of $\alpha = 4200\ s^{-1}$ were used to create the curve.

The bead pack acts as a low pass filter with a sharp cut-off at approximately 200kHz (blue curve in Fig. 2b). The signal is fairly narrowband with support mostly between 150-200kHz. It is strikingly different from the signal recorded on the upper plate (red curve in Fig. 2b), which has support up to and beyond 1MHz. Using the language of phononics, the spectrum can be thought of as the acoustical branch of an amorphous granular medium and consists of rigid body degrees of freedom for the beads under weak coupling [43].[2]

Following the work of Merkel *et al.* [44,45], a formula for the highest cutoff frequency of a hexagonal close-packed (hcp) granular crystal of rigid spheres can be obtained:

$$f_{cutoff} = \frac{1}{2\pi}\sqrt{\frac{40K_s}{m_b}} \quad (1)$$

where $m_b$ is the mass of a glass bead and $K_s$ is the shear rigidity of the inter-bead contacts. Of the three types of vibrational modes in a granular crystal—longitudinal, shear, and rotational—the rotational modes determine the highest cutoff frequency. Using Hertzian contact theory, the shear rigidity can be expressed in terms of material properties as:

$$K_s = (3aF)^{1/3}E^{2/3}\frac{(1-\nu^2)^{1/3}}{(1+\nu)(2-\nu)} \quad (2)$$

where $F$ is the force between beads, $\nu$ is the Poisson ratio, $E$ is the elastic modulus, and $a$ is the diameter of a bead. Estimating $F$ to be 1.5 N (from a 215kPa pressure and the cross-sectional area of a single 3mm bead)[3], and using $\nu = 0.22$, $E = 70$ GPa, $a = 3$mm, $m_b = 30$ mg, we obtain a cutoff frequency of $f_{cutoff} = 249$ kHz. This aligns adequately with the experimental data. The difference is ascribable to the bead pack corresponding better to random close pack than hexagonal, with lower coordination number than that of the perfect crystal. Furthermore, group velocity will be zero at the cutoff frequency, so one would not expect discernable amplitude at precisely $f_{cutoff}$.

The steel plates were chosen to be quite thin—1.6mm—to ensure that the resonances of the plates did not contaminate the spectrum and to enhance radiation loss rate into the beads. At 160kHz, the wavelength of a flexural wave in the steel plate is 10mm (the dispersion relation for flexural waves in a plate is $k = \sqrt{3.464\,\omega/c_{pl}h}$ where $c_{pl} = \sqrt{E_{steel}/(\rho_{steel}(1-\nu_{steel}^2))}$ and $h$ is the thickness of the plate). This wavelength is small compared to the plate diameter, so many modes exist in the plate, ensuring that the individual modes will not be well resolved. We further estimate that the lifetime of a flexural wave in the plate, against radiation into longitudinal waves in the bead pack, is short: $9\mu s (= \rho_{steel}h/\rho_{bp}c_{bp})$. This implies that a typical wave in the plate has travelled 27mm before entering the bead pack. Therefore, the intensity entering the bead pack is planar, as the ultrasound has spread across the steel plate before leaving it.

---

[2] We also observe the first two optical branches, at 900kHz and 1500kHz respectively (not shown). Transmission at these frequencies is very weak, however, and detection requires the use of a thinner (9mm compared to 33mm thick) bead pack.

[3] From the contact force of $F = 1.5N$, Johnson [49] tells us that radius of the contact circle is 36 microns, the maximum contact pressure is 559MPa, the maximum shear stress is 182MPa, and the maximum tensile stress is 104MPa. The tensile stress is just above the nominal tensile strengths in glass (10-100MPa). However, glass can have a tensile strength much higher. Microscope inspection of glass beads after disassembly did not reveal any cracks. Either there were none, or they closed upon disassembly. We suggest the use of tempered glass beads in the future to prevent cracking.

Foam is employed to support and surround the bead pack because it ensures that any leakage of ultrasound into the support will not return to the bead pack. Consequently, the coda wave interferometry analysis (next section) is not contaminated by waves that have spent time in the support structures rather than the beads. Separate measurements confirm that the foam has much greater absorptivity than the bead pack. Moreover, the impedance mismatch between the foam and beads is high: $Z_{bp}/Z_{foam} = 19$ (using the measured wave speed of 1250 m/s and density 59 kg/m³ for the foam). Thus, ultrasound that has explored the foam is not an important part of the signal received at the bottom (i.e., signals like Fig. 2a).

Because the walls of the bead pack are foam, there was concern that the slow dynamic conditioning and relaxation presented below (Sec. 4) are due to the foam itself rather than the bead pack. We replaced the foam walls with a brass cylinder and found the slow dynamics, though noisier, to have the same magnitude as with foam walls. Similarly, there was concern that the observed slow dynamic recovery was actually due to the rubber shims between the safety legs and the static load. However, we removed these shims, leaving an air gap, and found the slow dynamic magnitude to be unaffected, though the noise increased. Thus, we are confident that the slow dynamics results presented below are due to the glass bead pack and not other parts of our apparatus. We choose to present the results with the lower noise level.

### III. Coda Wave Interferometry

Coda wave interferometry (CWI) is used to quantify changes in the bead pack over time. Figure 4 summarizes the process. A normalized cross-correlation $X_n^i$ is constructed between a reference signal, $\phi$, and all signals produced in a measurement, $\psi_n$, which were captured at laboratory-times $T_n$. The cross-correlation is over a certain signal-time window $i$ centered at $t^i$ and having a width $W$:

$$X_n^i(\tau) = \frac{1}{A_n^i} \int_{t^i - W/2}^{t^i + W/2} dt' \, \phi(t') \psi_n(t' + \tau) \tag{3}$$

where $A_n^i$ is the normalization factor:

$$A_n^i = \sqrt{\int dt' \, \phi^2(t') \int dt' \, \psi_n^2(t' + \tau)} \tag{4}$$

The integrals in $A_n^i$ are over the same time region as those in the numerator of $X_n^i$. We distinguish between "laboratory-time" and "signal-time" to emphasize the different time scales involved. Laboratory-time $T_n$ ranges from seconds to minutes and its index $n$ goes from 1 to $N$, where $N$ is the total number of signals captured in a measurement. (It is labeled as "time" in Figs. 5-8, below.) Window times $t^i$ range from 100s of microseconds to milliseconds, and the index $i$ labels a window of signal-time. The range of $i$ varies with window width $W$ and how much signal is being examined. $t'$ is signal-time after the main bang of the pulser and ranges from 0 to 3ms. Typically, we choose $W = 200$ μs and extend $i$ to include the signal up to 2.5 ms. The first signal-time window begins at $t^1 - W/2 = 50$μs.

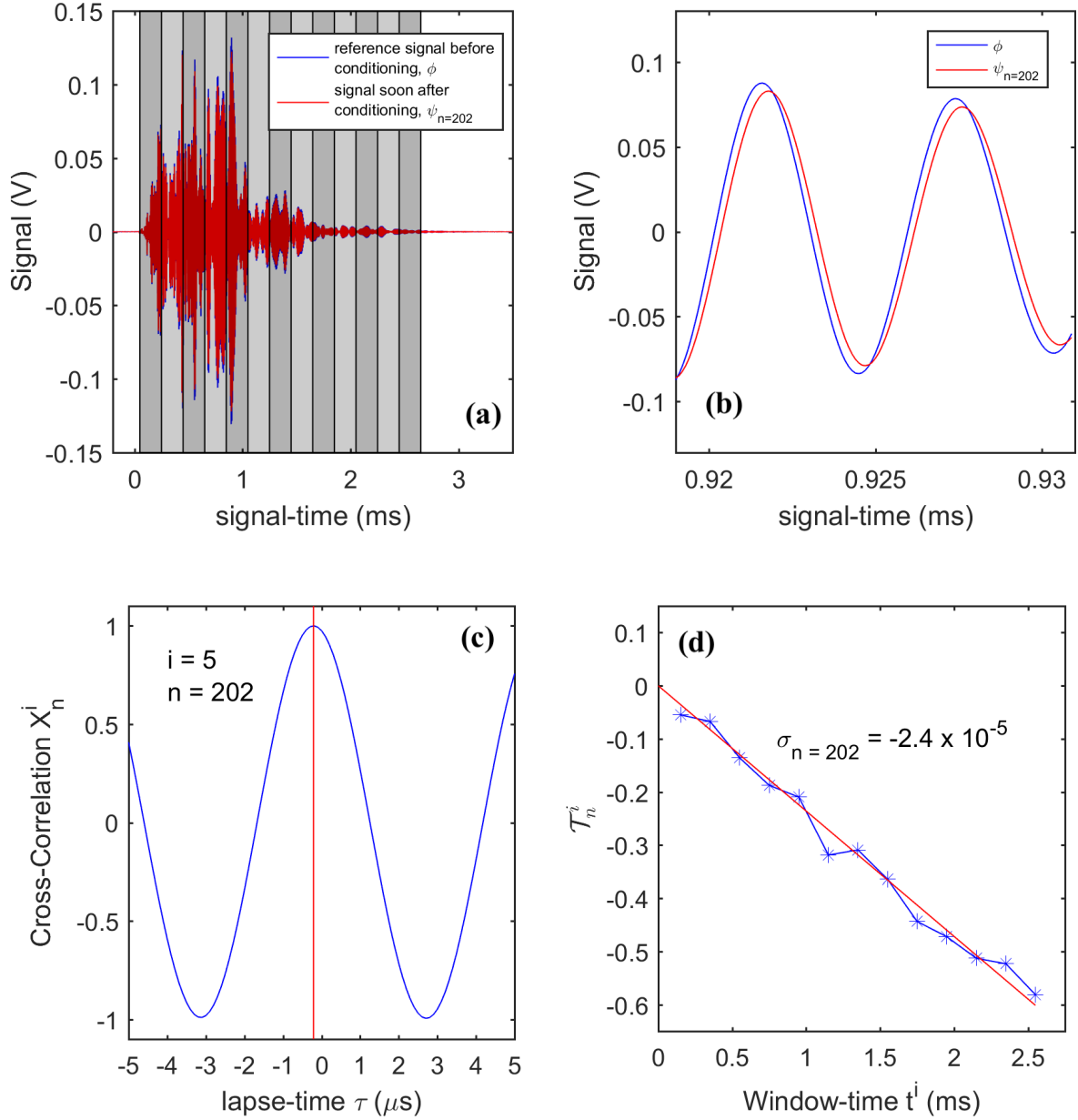

Figure 4. Summary of the coda wave interferometry procedure. Panel (a) shows the reference signal, $\phi = \psi_1$, (blue curve) and the 202$^{nd}$ signal (red curve) in a measurement. The plot is shaded to signify the 13 windows ($W = 200\ \mu s$) used to construct $X_{202}^{i=1,2,\ldots13}$. Panel (b) shows a magnified region of panel (a) to highlight that $\psi_{202}$ is delayed with respect to $\phi$. The delay is quantified by calculating the lapse-time value where $X_n^i$ is maximum, $\mathcal{T}_n^i = \arg\max X_n^i(\tau)$. Panel (c) shows $X_{202}^5(\tau)$ with its maximum at $\mathcal{T}_{202}^5$ designated by the red vertical line. The $\mathcal{T}_n^i$ values for a given $n$ are plotted versus signal-time $t^i$, as shown in the blue curve of panel (d) for $n = 202$. We fit the curve to a straight line (red curve) with zero y- intercept: $\mathcal{T}_n^i = \sigma_n t^i$. The slope $\sigma_n$ is called stretch. It is a signature of changes in the diffuse ultrasound and the sample.

We take the reference signal to be the first signal in a measurement: $\phi = \psi_1$, usually before conditioning is applied. Figure 4a plots an example comparison of two signals: the reference signal (blue curve) and the n=202$^{nd}$ signal (red curve), recorded about 600 seconds after $\phi$. Small differences exist between the signals. The plot is shaded to signify the 13 windows ($W = 200\ \mu s$) used to construct $X_{202}^{i=1,2,\ldots 13}$. Figure 4b shows an expanded region of Fig. 4a to highlight that $\psi_{202}$ is delayed with respect to $\phi$. The delay is quantified by calculating the lapse-time value where $X_n^i$ is maximum, $\mathcal{T}_n^i = \arg\max X_n^i(\tau)$. Figure 4c shows $X_{202}^5(\tau)$ with $\mathcal{T}_{202}^5$ designated by the red vertical line. The $\mathcal{T}_n^i$ values for a given $n$ are plotted versus signal-time $t^i$ (Fig. 4d, for $n = 202$). If the difference between the signals and the reference signal is a pure dilation, this plot should be linear with zero y-intercept [33,46]. Thus, we fit the ordered pairs $(t^1, \mathcal{T}_n^1), (t^2, \mathcal{T}_n^2) \ldots$ to a straight line:

$$\mathcal{T}_n^i = \sigma_n t^i \tag{5}$$

The slope $\sigma_n$ is called waveform dilation or "stretch." It can be interpreted as a relative change in wave speed and therefore a relative change in modulus: $\sigma = \frac{\Delta v}{v} = \frac{1}{2}\frac{\Delta M}{M}$. Hence, stretch will be our signature of changes in a sample. Observation of a log(lab-time $T_n$) recovery in stretch corresponds to log(time) recovery in the elastic modulus. With our sign conventions, a negative value of stretch means the signal $\psi_n$ is slower than the reference signal $\phi$.

The value of maximum correlation, $X_n^i(\tau = \mathcal{T}_n^i)$, is also potentially interesting but not pursued here. The logarithm of this value has been termed distortion and signifies differences between $\phi$ and $\psi_n$ that are not a simple dilation [33]. Something akin to distortion, called the "resemblance parameter" $X_n^i(\tau = 0)$, has been used previously in acoustic nonlinearity studies of glass bead packs [32,35]. It is an alternative way to quantify how the high frequency diffuse waveforms vary during conditioning and relaxation.

**IV. Slow Dynamics Results**

A pump-probe scheme is used to study slow dynamics in the bead pack, as used in NRUS experiments and elsewhere. Here the probe consists of the low-amplitude, noninvasive, multiply-scattered diffuse ultrasonic waves described above. As confirmed *aposteriori*, the probe ultrasound is of sufficiently low amplitude to ensure that the probe waves themselves are not significantly conditioning the bead pack.[4] For the CWI processing, a time window of $W = 200\ \mu s$ was used for all stretch calculations, and the first and last non-overlapping time windows were centered at $t^1 = 150 \mu s$ and $t^{13} = 2550 \mu s$, respectively.

Three methods of pumping are employed: impulsive, harmonic, and quasi-static (see Table 1). The first two were chosen to correspond to pump methods used by others [3,11,18,31]. The third, to our knowledge, has not been published in the literature to date.

---

[4] We confirmed that the ultrasound was not itself conditioning the bead pack by beginning a measurement with the pulse amplitude low. After a sufficient number of repetition-averaged pulses to establish a consistent value of stretch, the pulse amplitude was approximately quadrupled. The stretch values, after the pulse amplitude increased, were unchanged. We took this to be sufficient evidence that the ultrasound was not itself conditioning the bead pack. Taking extra care, we kept the pulse amplitude at the initial low value for all slow dynamic experiments shown here.

| Type of conditioning | Description | $\epsilon$ – Estimated strain | $m$ – Slope of sigma vs ln(T) plot |
|---|---|---|---|
| Impulsive | Dropped rubber ball from 0.2 meters | $\epsilon_{peak} = 2.8 \times 10^{-5}$ | $4.2 \times 10^{-5}$ |
| Harmonic | Dynamical shaker resting on top of static load (label (v) in Fig. 1b) | $\epsilon_{rms} = 6.15 \times 10^{-7}$ | $1.4 \times 10^{-5}$ (shaker off)<br><br>$-9.7 \times 10^{-6}$ (shaker on) |
| Quasi-static | Added and subtracted 1kg mass from 87kg static load | $\epsilon = 2.15 \times 10^{-6}$ | $2.4 \times 10^{-5}$ (adding weight)<br><br>$7.9 \times 10^{-6}$ (subtracting weight) |

Table 1. Summary of different conditioning used in the slow dynamics experiments (Sec. IV and Figs. 5-7) as well as the estimated strains associated with each conditioning and the slopes of the recovery. The slopes were estimated by fitting the recovery from 15 seconds after recovery to 3 minutes after recovery.

**A. Slow dynamics from impulsive conditioning**

Our impulsive pump is a rubber ball (mass of 150g, diameter of 6.25cm) dropped from 0.2m on top of the 87kg static load. Impulsive pumping has been previously used on cement paste and sandstone samples by dropping a small wooden ball [11] and on concrete samples by dropping a small metal ball [12]. Primary benefits of impulsive pumping are a clear time of conditioning and ease of application [11,12].

Results for impulsive conditioning of the glass bead pack are shown in Fig. 5. Slow dynamics is observed, as the characteristic drop in stretch followed by a slow recovery is clearly evident in Fig. 5a. The recovery is also clearly logarithmic in lab-time (Fig. 5b): $\sigma = m \ln(T_n - T_z) + $ constant, where the slope $m = 4.2 \times 10^{-5}$, and $T_z$ was chosen to give good linearity at early times.[5] This slope was estimated by fitting the recovery from 15 seconds to 3 minutes after impact. (This time period was used for the slope estimates below as well.) The time for full recovery, i.e. when the curve in Fig. 5a would cross the σ=0 axis, can be estimated as 50 hours. Observation of full recovery is, however, difficult due to potential contamination by drifts in temperature and/or ongoing slow recoveries after earlier mechanical disturbances. We do not attempt it here. The uptick in slope after 400 seconds is not meaningful; it could be ascribed to a temperature drift.

---

[5] We do not have independent measure of the zero time $T_z$ (equal to the time of impact, the time at which harmonic conditioning starts or ceases (Sec. IV.B), or the time at which the quasi-static loads are changed (Sec. IV.C)). The data itself, however, clearly indicate this zero time to within the three second interval between data points. In the plots (Figs. 5b, 6b, 7b, and 8a) we adjust $T_z$ so as to make them fully linear. Different choices for $T_z$ within the known interval will distort the linearity only for the first few data points. Strictly speaking, we only demonstrate linearity for $T_n - T_0 > \sim 15s$, but we note that good linearity has been observed back to 18msec [11]. It is recommended that future work record the zero time independently, e.g., [11]. The potential to unambiguously record the zero time is one of the advantages of impulsive conditioning.

The strain induced by the ball drop can be estimated using the formula:

$$\epsilon = \frac{F/A}{\rho c^2} = \frac{LF}{m_{bp} c^2} = (2.19 \times 10^{-7})F \tag{6}$$

where L = 33mm is the thickness of the bead pack, $m_{bp} = 221g$ is the mass of the pack, c = 825 m/s is the low-frequency wave speed, and $F$ is the force on the bead pack in Newtons. We estimate the impulsive force by placing an accelerometer on top of the load, which has a mass of $m_L = 87 kg$. Assuming the peak strain occurs at a time after the ball has rebounded[6], we identify the maximum acceleration 6 milliseconds after the impact: $a_L = 0.15g$ ($g = 9.81 m/s^2$ being standard gravity). Thus, the peak force is $F_{rb} = m_L a_L = 128N$, and we obtain a peak strain of $\epsilon_{peak} = 2.8 \times 10^{-5}$ using equation (6).

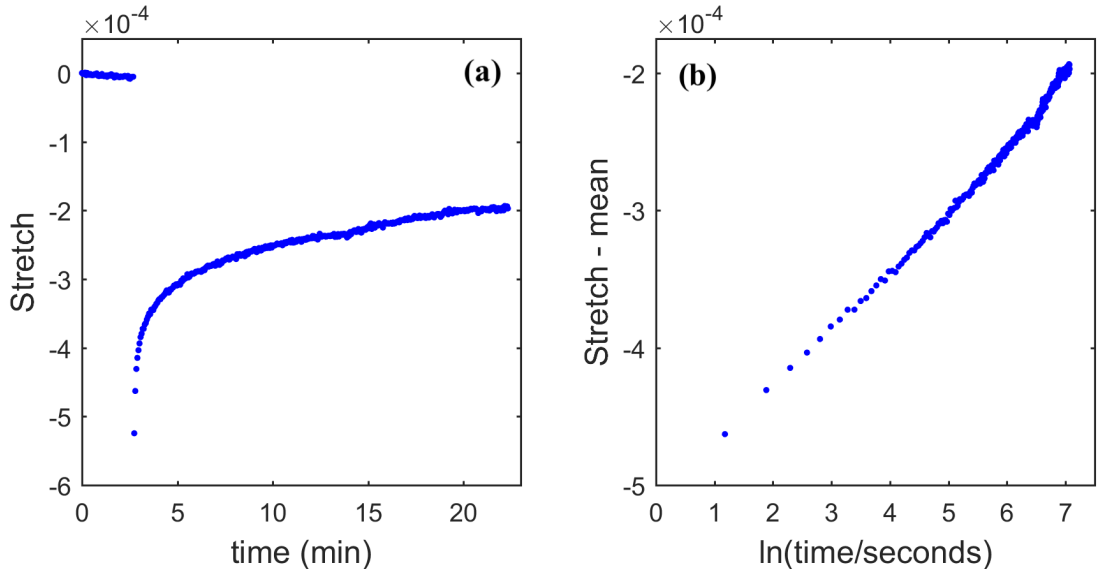

Figure 5. The slow dynamics results for impulsive pumping. Stretch is plotted versus laboratory-time in panel (a). The recovery is logarithmic in time, as seen in panel (b): stretch versus the logarithm of time after the ball drop at $T_{drop} \sim 3$ minutes. The slope (Table 1) was estimated by fitting the recovery from 15 seconds after recovery to 3 minutes after recovery.

---

[6] Hertzian contact theory provides an estimate for the contact time of the rubber ball with the static load. The contact time is given by $T^* = 2.87 \left(\frac{m_{rb}^2}{RE^2 v_i}\right)^{1/5} = 2.2\ ms$, where R is the radius of the rubber ball, $m_{rb}$ is the mass of the ball, $v_i = \sqrt{2gh_i} = 2m/s$ is the speed of the ball before impact, and $1/E = (1 - v_1^2)/E_1 + (1 - v_2^2)/E_2$ with $v_1, E_1$ ($v_2, E_2$) being the Poisson ratio and Young's modulus, respectively, of rubber (steel). We used 0.50 and 0.29 for the Poisson ratio of rubber and steel and 28MPa and 200GPa for the Young's modulus of rubber and steel.

## B. Slow dynamics from harmonic conditioning

The harmonic pump is a dynamical shaker, which rests on top of the static load and vibrates a 1kg mass (label (v) in Fig. 1b). The driving frequency was $f_D = 60Hz$, far from the fundamental longitudinal resonance of the structure ($f_{res} = 20Hz$). Much of the work from LANL [1–4,18,31] used harmonic pumping because NRUS was employed to measure slow dynamics. NRUS experiments used a sustained sinusoidal excitation at the longitudinal resonance of the sample for conditioning. One advantage of harmonic pumping is the ability to control and easily measure how much pump strain is being exerted on the sample.

The results for our harmonic conditioning are shown in Fig. 6. The shaded regions in Fig. 6a indicate when the shaker was on. There is extra noise in these regions associated with the shaker vibrations contaminating the ultrasonic signals. Again, slow dynamics is observed during the recovery (shaker off). Slow dynamics is also observed during the conditioning (shaker on), as the value of stretch drops suddenly and then more slowly continues to decrease. This is consistent with NRUS experiments (e.g., Fig. 2 in ref. [18]). Figure 6b shows that both the conditioning and recovery are logarithmic in time. The magnitude of the slopes are similar, but not the same (see Table 1). Determination of an extrapolated time for full recovery with harmonic conditioning is difficult because it is not clear to what value stretch is recovering. For impulsive conditioning, it was straight-forward: zero value of stretch. For harmonic conditioning, the quiescent state is distorted by previous cycles of conditioning and relaxation. However, some cycling is necessary, as the sample must first reach a steady state; one period of conditioning is not sufficient (Fig. 6a).

The strain induced by the shaker can be estimated by attaching an accelerometer to the 87kg static load. The force on the bead pack was calculated by

$$F_{har} = k_{eff} u(t) = (m_L \omega_{res}^2)(-a_L/\omega_D^2) = -m_L(f_{res}/f_D)^2 a_L(t) \quad (7)$$

where $m_L$ is the mass of the load, $u$ is the vertical displacement of the slabs, and $a_L$ is their acceleration. We use the rms of the accelerometer signal to determine $a_L$. For the measurement in Fig. 6, $a_L = 0.03g$, and the rms force on the bead pack is consequently $F_{rms} = 2.8N$. Using equation (6) above, we obtain a steady-state rms strain of $\epsilon_{rms} = 6.15 \times 10^{-7}$.

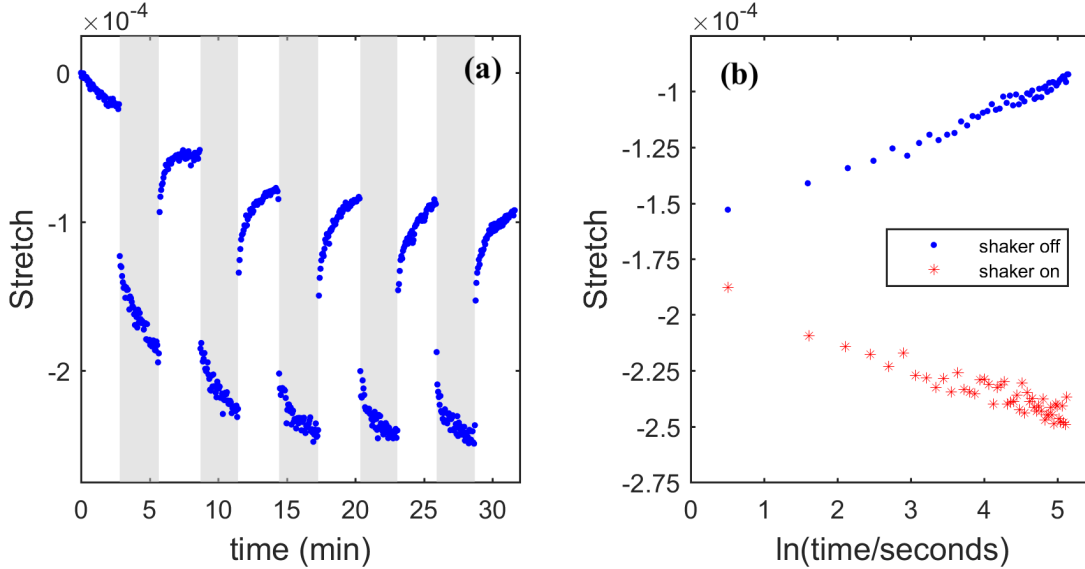

Figure 6. The slow dynamic results for harmonic pumping. Stretch is plotted versus laboratory-time in panel (a). The shaded regions indicate conditioning, i.e., the shaker was on in the shaded regions. Both conditioning and recovery are logarithmic in time. Panel (b) shows stretch versus the logarithm of time for the last conditioning and recovery sections. The slopes (Table 1) were estimated by fitting the recovery from 15 seconds after recovery to 3 minutes after recovery.

## C. Slow dynamics from quasi-static conditioning

The quasi-static pump involved the periodic adding and subtracting of an additional mass on top of the static load. The additional mass was 1kg. The results are shown in Fig. 7. The shaded region in Fig. 7a denote when the 1kg was added. It is expected that relative wave speed would increase when the extra mass was added (speed $\propto F^{1/6}$, according to Hertzian theory [36]) leading to a positive stretch. However, that the wave speed would continue to increase logarithmically (Fig. 7b) after the addition is a sign of nontrivial slow dynamics. Similarly, when subtracting the mass, we expect stretch to return to its initial value, i.e. zero. Rather, the measured value overshoots the expected value and then recovers slowly towards it. The strain for quasi-static conditioning can be estimated using equation (6) above, where now $F_{qs} = (1kg) \times (9.81 m/s^2) = 9.81N$. The strain from adding the extra mass is consequently $2.15 \times 10^{-6}$.

For quasi-static conditioning we can also predict how much stretch should occur long after adding the extra 1kg mass. By Hertzian theory, speed should be proportional to the sixth root of the force. The addition of 1kg increases the static force on bead pack approximately 1.1%, so the fully relaxed stretch should be $0.011/6 = 1.8 \times 10^{-3}$. Our measured stretch after three minutes of $3 \times 10^{-4}$ is less than this by a factor of 6. An extrapolation based on the observed slope, $m = 2.4 \times 10^{-5}$, indicates that it would take many times the age of the universe ($10^{21}$ years) to reach the predicted value of final stretch. This striking number, and its difference from the 50hr extrapolated time to full recovery for impulsive conditioning, begs to be explained. It cannot be ascribed to a mis-estimate of $T_z$.

Slow dynamic experiments with quasi-static conditioning have not been reported in the literature previously. The results show clearly the symmetry breaking of the inducing source, as both tensile and compressive conditionings lead to a relaxation characterized by a slow dynamic *increase* in modulus, regardless of the sign of the pumping. This asymmetry has been emphasized by TenCate *et al.* [2] as a key characteristic of slow dynamics and distinguishes it from other creep phenomena.

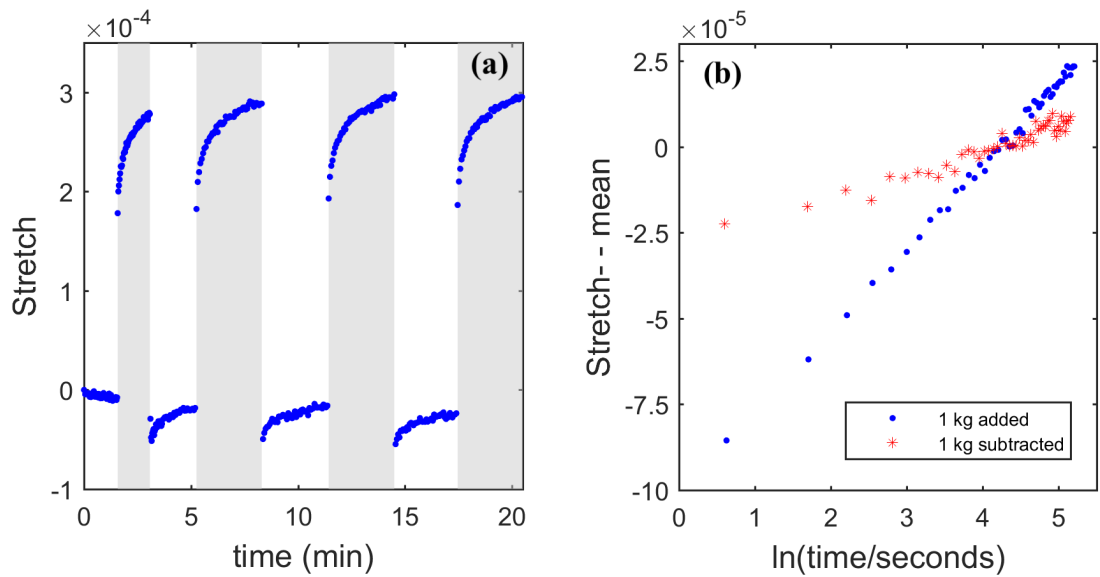

Figure 7. The slow dynamic results for quasi-static pumping. Stretch is plotted versus the laboratory-time in panel (a). The shaded regions indicate the times in which the 1kg mass was placed on top of the large static load. Both conditioning (1kg added) and recovery (1kg subtracted) are logarithmic in time. Panel (b) shows stretch minus its mean versus the logarithm of time for the last conditioning and recovery sections. The slopes (Table 1) were estimated by fitting the recovery from 15 seconds after recovery to 3 minutes after recovery. The mean is subtracted to more easily plot both curves in one panel.

## V. Discussion

Glass bead packs offer three advantages over the more commonly used materials to study slow dynamics. First, bead packs have a simplified chemistry and history compared to these materials. Where cement pastes need many days to cure and sandstones are created over the course of millions of years. Glass bead packs, on the other hand, have virtually no history. Second, bead packs have many internal surfaces that are easier to characterize than the grain contacts present in sandstones or cement. Third, glass bead packs can have a high porosity and large pores, enabling better control of the environment at the contact points. It can be difficult to control, or even know, the internal structure and environment within cement paste or sandstones.

Here we have not attempted to test proposed mechanisms for slow dynamics. However, we do argue that unconsolidated glass bead packs may provide a useful experimental venue in which to study slow dynamics, and that they are particularly suited for determining how slow dynamic behavior may depend on i) environmental factors, like humidity and temperature, ii) the properties of the bead pack, like grain size, pack thickness, material, and bead surface treatments, and iii) changes to the medium, like saturating the bead pack with water or other liquids. Here we have shown that ultrasonic probes and CWI processing allow for great precision in measuring stretch in glass bead packs. Precision for the stretch measurements here appears to be better than $10^{-6}$. The noise level in the slow dynamic measurements reported here (Figs. 5-7) is lower than previous slow dynamic experiments in these media (e.g., Fig. 2c in ref. [31]).

It has been shown that this bead pack system exhibits slow dynamics—when probed by ultrasound combined with CWI and for a variety of pump methods—and does so with good precision. We conclude with an application: a preliminary study of the slow dynamic recovery dependence on rms pump strain. TenCate *et al.* [2] showed that recovery slope $m$ is linear in rms pump strain, at least for $\epsilon^{rms} > 10^{-6}$. Below that, $m$ levels off. Their minimum and maximum strain values used were $0.40 \times 10^{-6}$ and $2.64 \times 10^{-6}$. Here we repeat the measurement of Sec. IV.B and Fig. 6, for rms strains of $\epsilon^{rms} = 5.74 \times 10^{-8}$, $1.13 \times 10^{-7}$, $2.72 \times 10^{-7}$, $6.15 \times 10^{-7}$ and $1.32 \times 10^{-6}$. The results are shown in Fig. 8. Like TenCate *et al.* [2], we see an apparent linear regime at pump strains near one microstrain and a leveling off below that (Fig. 8b). Attempts to measure slow dynamics with pump strains lower than $5.74 \times 10^{-8}$ were unsuccessful as the response was contaminated by drifts, presumably from temperature changes.[7] If temperature were controlled, it might be possible to investigate if slow dynamics persists at even lower pump strains.

Other investigations [47,48] of general nonlinear and nonequilibrium behavior in rocks (i.e., not confined to slow dynamics) have identified a threshold strain, $\epsilon_M$, below which the nonequilibrium nonlinear behavior, including slow dynamic nonlinearity, does not occur and above which it does. For Berea sandstone $\epsilon_M = 5 \times 10^{-7}$, while for Fontainebleau sandstone $\epsilon_M = 2 \times 10^{-7}$ [47]. The threshold was calculated by first measuring the resonant frequency $f_0$ of the sample (the fundamental longitudinal mode in a rod of the material) at very low strain ($\sim 10^{-9}$), second driving the sample at a higher strain $\epsilon_D$, and third measuring $f_0$ again. By repeating this three-step process for different values of $\epsilon_D$, the threshold could be determined. For $\epsilon_D < \epsilon_M$, $f_0$ did not change when it was measured after the sample was driven at $\epsilon_D$; for $\epsilon_D > \epsilon_M$, $f_0$ did change (and would subsequently relax back logarithmically in time to its original value taken before the conditioning at $\epsilon_D$).

---

[7] While temperature drifts are a plausible explanation for the slight accelerations in recoveries in Fig. 8, it remains only an hypothesis.

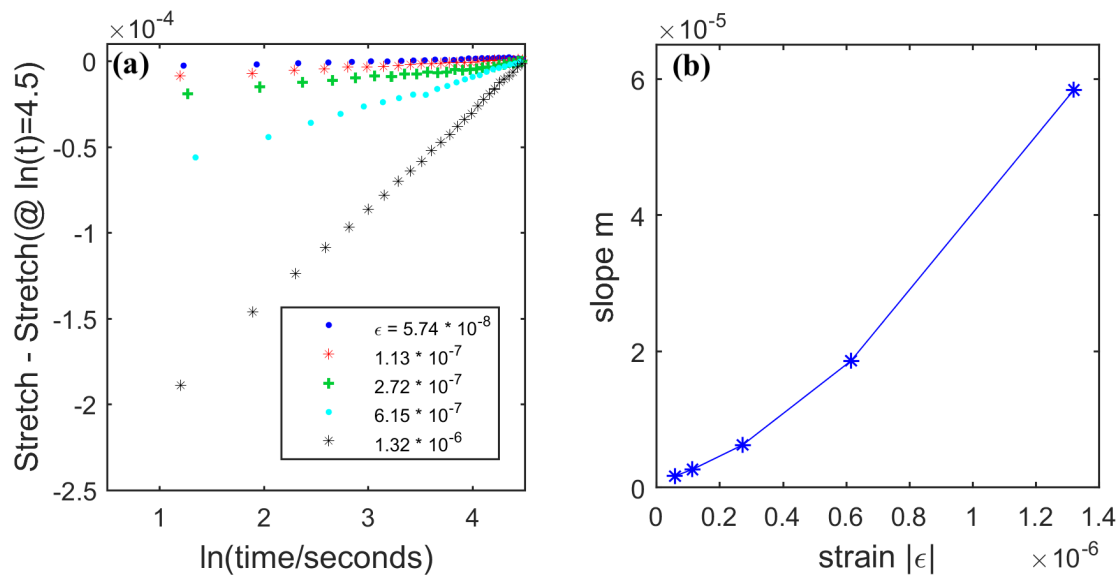

Figure 8. Slow dynamics recovery for different pump strain levels. Panel (a) shows stretch versus log(time) for the last recovery after five cycles of harmonic conditioning and relaxation (i.e., the shaker was turned on and off five times). The voltage of the shaker was adjusted to change the strain level. The method for estimating strain level is presented in Sec. IV.B. Panel (b) shows the early time slope of the recovery in panel (a) versus strain level. The slopes were estimated by fitting the recovery from 15 seconds after recovery to 3 minutes after recovery. There is a linear relationship between slope and strain, for strains at and above 1 microstrain, as observed elsewhere [2].

Johnson and Jia [31] attempted to determine a threshold strain for glass bead packs. They stated that $\epsilon_M$ should be of order a microstrain, though it will increase with pressure. However, the results shown here (Fig. 8) indicate that the threshold may be much lower—below $6 \times 10^{-8}$—if it exists at all, as we still observe log(t) recovery (i.e. nonequilibrium behavior) at this strain level. (The pressure in our measurements, 215kPa, falls within the range of Johnson and Jia, 70 – 300kPa.) We emphasize here that "threshold strain" is defined as the strain below which nonequilibrium behavior does *not* occur [31,47,48]. This may not be the most useful definition. Rather, we suggest defining the threshold strain as the strain below which there is no longer a linear relationship between strain and the slope $m$ of the log(t) recovery. Strain values below this alternative threshold would still incur slow dynamics. The leveling off observed here (Fig. 8b) indicates that there is no clear delineation between two strain regimes (where slow dynamics occurs and where it does not). This may have implications for dynamical earthquake triggering, for the strain threshold was emphasized by Johnson and Jia [31] as being part of the triggering mechanism. The absence of a sharp delineation is also significant for determining if a relationship exists between hysteresis and slow dynamics, as it is widely believed that the threshold pertains to both behaviors [21,22].

Moreover, the paradigm of associating slow dynamics with the changes in the resonant frequency $f_0$ of bar experiments may not be the most helpful. Even though $f_0$ remains unchanged, it is not guaranteed that slow dynamic nonlinearity is not still occurring. Ultrasonic waves (100s of kHz) are more sensitive to changes in a sample than the resonant frequencies (1-10kHz), and CWI takes advantage of this sensitivity. Work with DAET [7] and others [11,12] have already demonstrated the value of ultrasonic waves. It would be of interest to employ ultrasound and CWI on Berea and Fontainebleau sandstone samples with pump strains lower than $\epsilon_M$ to determine if slow dynamics remains.

## VI. Conclusion

In this paper, we have presented an alternative experimental venue in which to study the poorly understood nonlinear elastic phenomena of slow dynamics. The material used here is unconsolidated glass bead packs, which in themselves offer advantages over other materials for slow dynamics studies. Our careful experimental design, which includes floating walls, a static dead-weight load, foam surroundings, and the use of ultrasonic wave probes with CWI processing provides low noise and great sensitivity to changes in the bead pack. The combination of these constituents has provided clear observation of slow dynamic relaxation. We have also demonstrated slow dynamic response to a variety of low frequency pumps—not just harmonic but also impulsive and quasi-static pumping. We anticipate future methodical tests of slow dynamic dependence on sundry parameters.

**Acknowledgements**

We are most grateful to John Popovics and James Bittner for their interest and encouragement, for many fruitful discussions, and their advice with the early stages of the experimental design.